\documentclass[11pt]{article}
            
\usepackage{fullpage}
\usepackage[numbers]{natbib}
\usepackage[auth-sc-lg]{authblk}

\usepackage{times}
\usepackage{soul}
\usepackage{url}
\usepackage[hidelinks]{hyperref}
\usepackage[utf8]{inputenc}
\usepackage[small]{caption}
\usepackage{graphicx}
\usepackage{amsmath}
\usepackage{mathrsfs}
\usepackage{amsthm}
\usepackage{amssymb}
\usepackage{booktabs}
\usepackage{algorithm}
\usepackage[noend]{algpseudocode}

\usepackage[shortlabels]{enumitem}
\usepackage[capitalize]{cleveref}
\usepackage{xspace}
\usepackage{bm}
\usepackage[f]{esvect}

\usepackage{tikz}
\usetikzlibrary{arrows}

\DeclareMathSymbol{\R}{\mathord}{AMSb}{"52}

\newcommand{\name}{\textsc{PluralityVeto}\xspace}
\newcommand{\randomized}[1]{$#1$-\textsc{RoundPluralityVeto}\xspace}
\newcommand{\fractionalveto}{\textsc{FractionalVeto}\xspace}
\newcommand{\randomdictatorship}{\textsc{RandomDictatorship}\xspace}
\newcommand{\pluralitymatching}{\textsc{PluralityMatching}\xspace}

\newcommand{\elec}{\mathcal{E}}
\newcommand{\prof}{{\vv{\succ}}}
\renewcommand{\top}{\mathsf{top}}
\renewcommand{\bot}{\mathsf{bottom}}
\newcommand{\plu}{\mathsf{plu}}
\newcommand{\dom}{\mathcal{D}}
\newcommand{\cost}{\mathsf{cost}}
\newcommand{\dist}{\mathsf{dist}}
\newcommand{\score}{\mathsf{score}}
\newcommand{\weight}{\mathsf{weight}}

\newcommand{\pvec}[1][]{\ifthenelse{\equal{#1}{}}{\mathbf{p}}{\mathbf{p}^{#1}}}
\newcommand{\qvec}[1][]{\ifthenelse{\equal{#1}{}}{\mathbf{q}}{\mathbf{q}^{#1}}}
\newcommand{\p}[2][]{\ifthenelse{\equal{#1}{}}{p_{#2}}{p^{#1}_{#2}}}
\newcommand{\q}[2][]{\ifthenelse{\equal{#1}{}}{q_{#2}}{q^{#1}_{#2}}}
\newcommand{\wvec}{\mathbf{w}}
\newcommand{\w}[1]{w_{#1}}
\newcommand{\f}{g}
\newcommand{\barV}{\xbar{V}}
\newcommand*\xbar[1]{%
  \hbox{%
    \vbox{%
      \hrule height 0.5pt 
      \kern0.3ex
      \hbox{%
        \kern-0.1em
        \ensuremath{#1}%
        \kern-0.1em
      }%
    }%
  }%
}
\newcommand{\barVp}{\ybar{V'}}
\newcommand*\ybar[1]{%
  \hbox{%
    \vbox{%
      \hrule height 0.5pt 
      \kern0.2ex
      \hbox{%
        \kern-0.1em
        \ensuremath{#1}%
        \kern-0.33em
      }%
    }%
  }%
}

\DeclareMathOperator*{\argmin}{arg\,min}
\renewcommand{\succeq}{\succcurlyeq}

\newtheorem{lemma}{Lemma}
\newtheorem{theorem}{Theorem}
\newtheorem{result}{Contribution}

\newcommand{\LPlabel}{}

\newenvironment{LP}[3][]{%
\renewcommand{\LPlabel}{#1}
\ifthenelse{\equal{\LPlabel}{}}{%
\[ \begin{array}{ll}
\mbox{#2} & \;\;#3 \\
\mbox{subject to} & \begin{array}[t]{ll}%
}{%
\begin{equation} \begin{array}{ll}
\mbox{#2} & \;\;#3 \\
\mbox{subject to} & \begin{array}[t]{ll}%
}}{%
\ifthenelse{\equal{\LPlabel}{}}{%
\end{array} \end{array} \]}{%
\end{array} \end{array} \label{\LPlabel} \end{equation}}%
}

\newenvironment{rtheorem}[3][]{

\bigskip

\noindent \ifthenelse{\equal{#1}{}}{\bf #2 #3}{\bf #2 #3 (#1)}
\begin{it}
}{\end{it}}

\providecommand{\SET}[1]{\ensuremath{\{ #1 \}}\xspace}
\providecommand{\Set}[2]{\ensuremath{\SET{#1 \mid #2}}\xspace}

\newcommand{\Pref}[3][]{\ensuremath{#2 \succ_{#1} #3}\xspace}
\newcommand{\LPC}{\ensuremath{c^*}\xspace}
\newcommand{\Lpm}[2]{\ensuremath{x_{#2,#1}}\xspace}
\newcommand{\DNo}{\ensuremath{\alpha}\xspace}
\newcommand{\DCo}[3]{\ensuremath{\phi^{(#1)}_{#2,#3}}\xspace}
\newcommand{\DTe}[4]{\ensuremath{\psi^{(#1,#2)}_{#3,#4}}\xspace}

\title{\name: A Simple Voting Rule \\ Achieving Optimal Metric Distortion}

\author[ ]{Fatih Erdem Kizilkaya\thanks{Contact Author}}
\author[ ]{ David Kempe}

\affil[ ]{\textit{University of Southern California}} 
\affil[ ]{\normalsize \tt \{Fatih.Erdem.Kizilkaya, David.M.Kempe\}@gmail.com}

\date{}

\begin{document}

\maketitle

\begin{abstract}
The metric distortion framework posits that $n$ voters and $m$ candidates are jointly embedded in a~metric space such that voters rank candidates that are closer to them higher.
A voting rule's purpose is to pick a candidate with minimum total distance to the voters, given only the rankings, but not the actual distances.
As a result, in the worst case, each deterministic rule picks a candidate whose total distance is at least three times larger than that of an optimal one, i.e., has distortion at least 3.
A recent breakthrough result showed that achieving this bound of 3 is possible;
however, the proof is non-constructive, and the~voting rule itself is a complicated exhaustive search.

Our main result is an extremely simple voting rule, called \name, which achieves the same optimal distortion of 3. 
Each candidate starts with a~score equal to his number of first-place votes.
These scores are then gradually decreased via an $n$-round veto process in which a candidate drops out when his score reaches zero. One after the other, voters decrement the score of their bottom choice among the standing candidates, and the last standing candidate wins.
We give a one-paragraph proof that this voting rule achieves distortion 3.
This rule is also immensely practical, and it only makes two queries to each voter, so it has low communication overhead.
We also show that a straightforward extension can be used to give a constructive proof of the more general Ranking-Matching Lemma of Gkatzelis et al.

We also generalize \name into a class of randomized voting rules in the following way:
\name is run only for $k < n$ rounds; then, a candidate is chosen with probability proportional to his residual score.
This general rule interpolates between \randomdictatorship (for $k=0$) and \name (for $k=n-1$), and $k$ controls the variance of the output.
We show that for all $k$, this rule has expected distortion at most 3.
\end{abstract}

\section{Introduction}
\label{sec:intro}
Voting is a fundamental process for reaching consensus and plays a vital role in democracies, organizations and businesses. 
Even honeybees use a type of voting to decide among potential nest sites; each bee casts a \emph{numerical} vote indicating the intensity of its preferences, and the site with the highest score wins \cite{HoneybeeDemocracy}. 
Quantifying preference intensity is not so easy for complicated problems that humans face;
thus, the predominant approach is to elicit, from each voter, a preference \emph{ranking} over candidates.
This creates a need for a \emph{voting rule} that determines the winner from these preferences.
Numerous rules have been proposed over the years, with no consensus on a ``best'' rule.

A key contribution of computer science in this regard has been viewing the ranking of a voter as a proxy for the latent cost she\footnote{For ease of presentation, we use female pronouns for voters and male pronouns for candidates throughout.} incurs if some candidate wins. 
Then, an optimal candidate can be defined as one minimizing the total cost,
and vote aggregation can be interpreted as an optimization problem with missing information.
Due to the missing information, a voting rule can be thought of as an approximation algorithm, whose worst-case performance is referred to as its \emph{distortion} in this setting.

Without any structures on the costs, not much can be achieved in terms of distortion \cite{Caragiannis2017}.
An important contribution due to \citet{anshelevich:bhardwaj:postl}
(see also the journal version \citep{anshelevich:bhardwaj:elkind:postl:skowron} and recent surveys \citep{anshelevich:filos-ratsikas:shah:voudouris:retrospective,anshelevich:filos-ratsikas:shah:voudouris:reading-list}) was to assume that the $n$ voters and $m$ candidates are jointly embedded in a metric space, and voters rank candidates by increasing distance.\footnote{This assumption generalizes the classic notion of \emph{single-peaked preferences} \cite{black:rationale,moulin:single-peak}.}
This viewpoint is motivated by observing that each candidate exhibits a~standpoint on various issues that voters care about, 
and each voter also has a standpoint on these issues that is presumably reflected on her ballot.
One would then expect that voters rank candidates whose standpoints are ``closer'' to theirs higher.
Note that voting rules do not have access to this space; the only available information is the rankings, which serve as an \emph{ordinal} proxy for the (cardinal) distances. 
The worst-case approximation specifically for metric costs is called \emph{metric distortion}. (Formal definitions of all concepts are given in \cref{sec:model}.)

The metric distortion framework has proved to be a fruitful analysis tool. 
In their initial work, \citet{anshelevich:bhardwaj:postl} established a lower bound of 3 on the distortion of any deterministic voting rule, and showed that the Copeland rule nearly matches the lower bound by achieving distortion~5.
Several subsequent papers worked towards closing this gap.
Initially, the Ranked Pairs rule was conjectured to achieve distortion~3.
This was disproved by \citet{goel:krishnaswamy:munagala} who gave a lower bound of 5; \citet{DistortionDuality} strengthened the lower bound to $\Omega(\sqrt{m})$.
The first improvement over the upper bound of 5 was due to \citet{munagala:wang:improved}, who achieved distortion $2+\sqrt{5} \approx 4.23$ using a novel asymmetric variant of the Copeland rule.
The distortion-3 conjecture was recently resolved in a breakthrough result by \citet{gkatzelis:halpern:shah:resolving}, using a~novel voting rule called \pluralitymatching.

One of the main drawbacks of \pluralitymatching is that it is unusually complex for a voting rule in the conventional sense.
The winner is selected based on perfect matchings in certain bipartite graphs, which we will discuss shortly.
Due to the complex nature of the voting rule, it is not even obvious that there always exists a winner in \pluralitymatching; indeed, this existence proof was the main contribution of \citet{gkatzelis:halpern:shah:resolving} over the prior work of \citet{munagala:wang:improved} and \citet{DistortionDuality}.
The rule is also almost certainly too technical to be understood by the general public.

\bigskip

Our main contribution is an extremely simple voting rule, called \name, which achieves the~same optimal metric distortion of 3.  

\bigskip

Under \name, each candidate starts with a score equal to his \textit{plurality score}, i.e., the number of first-place votes he receives. 
These scores are then gradually decreased; when the score of a candidate reaches zero, he is eliminated. 
Voters are processed one by one in an arbitrary order: when it is the turn of a voter, she decrements the score of her bottom choice among uneliminated candidates.
Since the initial scores add up to the number of voters, all of the candidates will be eliminated at the end.
The last eliminated candidate wins.
Notice that this rule does not even require access to the voters' full rankings.
Aside from the top choices, it only requires from each voter her bottom choice among uneliminated candidates.
Thus, when implemented via sequential queries to voters, \name also has low communication overhead.
We summarize our main result as follows.
(A formal presentation and proof are given in \cref{sec:distortion}.)

\begin{result}
\name has the optimal metric distortion of 3 and can be implemented to require each voter to communicate only $O(\log m)$ bits to the voting rule. 
\end{result}

A candidate with a strict majority of first-place votes wins under \name, regardless of the order in which voters are processed.
Thus, one can think of the elimination process (or, as we call it, \emph{multi-round veto}) as a runoff stage.
This makes our rule conceptually simple as well; it is just plurality voting followed by multi-round veto.
In this respect, it resembles instant runoff voting, which is used in national elections in several countries.
As in instant runoff voting, \name can be used by eliciting from each voter her full ranking, so that the runoff stage can be run instantly. 
Alternatively, it can be arranged as a two-stage election in which voters first cast a vote for their top choice, and in the following stage, each voter cancels the vote of another voter.

\pluralitymatching and its analysis are based on \emph{domination graphs}, a family of bipartite graphs $G(c)$ (one per candidate $c$) between voters and voters defined in \cref{sec:model}.
\citet{munagala:wang:improved} and \citet{DistortionDuality}
had shown that if $G(c)$ has a perfect matching, then $c$ has distortion at most 3.
The key contribution of \citet{gkatzelis:halpern:shah:resolving} was to show that such a candidate $c$ always exists.
In fact, they proved a generalization, called the Ranking-Matching Lemma, which shows the existence of a graph with a \emph{weighted} perfect bipartite matching in a more general class of bipartite graphs.
We also give (in \cref{sec:ranking-matching}) a constructive one-paragraph proof of this more general Ranking-Matching Lemma, using a variant \fractionalveto of \name which decreases weights fractionally, rather than integrally, in each step.

Then, we generalize \name to a class of randomized voting rules that choose a candidate with probability proportional to his residual score at the $k^{\text{th}}$ round of \name, which we refer to as \randomized{k}.
When $k=0$, this more general rule specializes to the well-known rule \randomdictatorship, which chooses the top choice of a uniformly random voter.
Hence, randomizing the outcome proportional to the scores achieves distortion $3-2/n$ when $k=0$, as shown in \cite{anshelevich:postl:randomized}.  
In \cref{sec:randomized}, we show that \randomized{k} in fact achieves distortion at most 3 for \emph{all} $k$. 
In that way, it interpolates between \randomdictatorship (for $k=0$) and \name (for $k=n-1$), and the parameter $k$ controls the variance of the output, but does not affect the distortion.
The intuition behind this is that \name repeatedly decreases scores for ``extreme'' candidates, and thus is likely to end up with central ones. We elaborate on this intuition in the context of Peer Selection in \cref{sec:conclusion}.

\begin{result}
\randomized{k} has expected metric distortion at most 3 for any number of rounds $k$.
\end{result}

This upper bound is almost tight, since any rule that can only elect candidates who are the top choice of at least one voter must have distortion at least $3-o(1)$ \cite{gkatzelis:halpern:shah:resolving}.
Also, since $k$ controls the variance of the output, at a high level, this result relates to the work of \citet{fain:goel:munagala:prabhu:referee} who are not only interested in the expected distortion of rules, but also in the expected \emph{squared} distortion, essentially forcing randomized rules to have lower variance in their distortion.

Lastly, we turn our attention to \emph{multi-winner voting rules} electing a committee of size $k > 1$, under an objective function defined in \cref{sec:committee}.
Here, as a direct corollary of our analysis of \name, we resolve the main open question of \citet{multiwinner}.
\citet{multiwinner} had given a multi-winner voting rule of distortion 3 running in exponential time, as well as a multi-winner voting rule of distortion 9 running in polynomial time, but left open the existence of a rule with distortion 3 that runs in polynomial time.
We~resolve this open question positively by combining \name with a reduction from \citet{multiwinner}.

\begin{result}
A multi-winner voting rule that adapts \name, achieves distortion 3, and runs in polynomial time.  
\end{result}

\paragraph{Other Related Work}
The utilitarian analysis of voting rules through the lens of approximation algorithms was first suggested in \cite{BCHLPS:utilitarian:distortion,caragiannis:procaccia:voting,procaccia:approximation:gibbard,procaccia:rosenschein:distortion}.
\citet{boutilier:rosenschein:incomplete,anshelevich:bhardwaj:postl} were the first to clearly articulate the tension between the objective of maximizing utility (or minimizing cost) and the available information, which is only ordinal; they also termed the resulting gap \emph{distortion}.
In the earlier work, such as \cite{BCHLPS:utilitarian:distortion,caragiannis:procaccia:voting,procaccia:approximation:gibbard,procaccia:rosenschein:distortion}, the focus was on (positive) utilities, and no additional assumptions (such as metric costs) were placed on the utilities.

The role of randomization in reducing distortion has been studied in several prior papers.
The fact that randomized voting rules can achieve expected distortion lower than 3 (the known lower bound for deterministic voting rules) was first shown by \citet{anshelevich:postl:randomized}, who showed that 
\randomdictatorship achieves expected distortion $3-\frac{2}{n}$. A slightly improved distortion of $3-\frac{2}{m}$ was achieved in \cite{DistortionCommunication} by randomizing between \randomdictatorship and \textsc{ProportionalToSquares}.
This upper bound is best possible among rules that only have access to each candidate's plurality score \cite{gross:anshelevich:xia:agree} --- in fact, \citet{gross:anshelevich:xia:agree} prove a more general lower bound, which establishes that when each voter only communicates her top $k < n/2$ candidates, every randomized voting rule has distortion at least $3-\frac{2}{\lfloor n/k \rfloor}$.
A lower bound of 2 on the distortion of any randomized voting rule is straightforward, and it had been conjectured that this bound may be achievable by some randomized voting rule.
This conjecture was recently disproved in \cite{charikar:ramakrishnan:randomized-distortion}, which established a lower bound of 2.0261 for $m=3$ candidates and 2.1126 as the number of candidates $m \to \infty$.
Whether any randomized voting rule can achieve expected disortion $3-\Omega(1)$ for arbitrary $m$ remains an intriguing open question.

Recall that \name only requires very limited communication from each voter, albeit in an $n$-round sequential algorithm.
In this way, our work relates generally to the study of communication in social choice rules 
(e.g., \cite{boutilier:rosenschein:incomplete,conitzer:sandholm:vote-elicitation}), and more specifically to studies of the tradeoff between communication and metric distortion.
The recent papers \cite{fain:goel:munagala:prabhu:referee,DistortionCommunication} establish related lower bounds: \citet{fain:goel:munagala:prabhu:referee} show that any voting rule that only obtains the top $k = O(1)$ candidates of each voter must have squared distortion $\Omega(m)$, in particular implying a bound of $\Omega(m)$ for the distortion of deterministic rules.
\citet{DistortionCommunication} proves a slightly more general and stronger lower bound of $\Omega(m/k)$ on the distortion of any deterministic voting rule that only obtains the candidates ranked by each voter in a set $K$ of size $k = |K|$ of positions.
Our voting rule avoids these lower bounds by obtaining the bottom candidate from a specified set for each voter; thus, for different voters, the candidates in different positions are queried.

Using randomization, communication can be drastically reduced even compared to our voting rule.
\citet{fain:goel:munagala:prabhu:referee} present a \textsc{RandomReferee} mechanism: the mechanism asks two randomly chosen voters for their top choices, and then has a third voter choose between the two proposed candidates. This mechanism, which only requires access to the top choices of two voters plus one bit, achieves not only constant expected distortion, but constant expected \emph{squared} distortion.

Several other recent works have studied the tradeoff between communication and distortion.
\citet{mandal:procaccia:shah:woodruff} study tradeoffs between communication and distortion in the utilitarian model, i.e., without any metric constraints. They also assume that voters actually know their utilities.
In this model, they obtain upper and lower bounds on the achievable distortion under communication complexity constraints.
\citet{pierczynski:skowron:approval} consider the distortion (and a modified notion of distortion) for approval-based voting (which has reduced communication), in which voters approve all candidates within a certain distance of themselves. They show that under certain parameter settings, for a carefully chosen radius, approval-based voting achieves constant distortion in their sense.
\citet{bentert:skowron:few-candidates} consider the approximate implementation of score-based voting rules using low communication. In particular, their techniques in Section~3.2 show that constant distortion $5+o(1)$ can be achieved when the number of voters is large, by asking each voter to compare two uniformly random candidates.

\section{Preliminaries}
\label{sec:model}
Throughout, we use bold face for vectors, and denote the $i^{\text{th}}$ component of a vector $\mathbf{x}$ by $x_i$.
Given a set $S$, let $\Delta(S)$ denote the probability simplex over $S$, i.e., the set of non-negative weight vectors over $S$ that add up to $1$.

An \emph{election} is a tuple $\elec = (V, C, \prof)$ consisting of a set of $n$ \emph{voters} $V$, a set of $m$ \emph{candidates} $C$ and a~\emph{ranked-choice profile} $\prof \ = (\succ_v)_{v \in V}$; here, $\succ_v$ is the \emph{ranking} of voter $v$, i.e., a total order over $C$.
We~say that voter $v$ ranks candidate $c$ \emph{higher} than candidate $c'$ if $c \succ_v c'$. 
We also use $c \succeq_v c'$ when  $c \succ_v c'$ or $c = c'$, in which case we say that $v$ ranks $c$ \emph{weakly higher} than $c'$. 

A \emph{voting rule} $f$ is an algorithm that returns a candidate $f(\prof) \in C$ given a ranked-choice profile $\prof$. We refer to $f(\prof)$ as the \emph{winner} of the election $\elec$ using the voting rule $f$, or just as the winner of $f$ if $\elec$ is clear from the context. 
For the most part, we will consider deterministic voting rules; in \cref{sec:randomized}, we will also study randomized voting rules.

We refer to the candidate ranked highest by voter $v$ as the \emph{top choice} of $v$, and denote him by $\top(v)$. The candidate ranked lowest by voter $v$ is likewise referred to as the \emph{bottom choice} of $v$.
We use $\plu(c)$ to denote the \emph{plurality score} of candidate $c$, i.e., the number of voters whose top choice is $c$.  

\subsection{Metric Distortion}
\label{sec:model:distortion}

A \textit{metric} over a set $S$ is a function $d : S \times S \rightarrow \mathbb{R}_{\geq 0}$ which satisfies the following conditions for all $a, b, c \in S$: 
(1) Positive Definiteness: $d(a,b) = 0$ if and only\footnote{Our proofs do not require the ``only if'' condition, so technically, all our results hold for pseudo-metrics, not just metrics.}  if $a=b$; 
(2) Symmetry: $d(a, b) = d(b, a)$; 
(3) Triangle inequality: $d(a,b) + d(b,c) \geq d(a,c)$.
  
Given an election $\elec = (V, C, \prof)$, we say that a metric $d$ over\footnote{We only care about the distances between voters and candidates, so $d$ can be defined as a function $d: V \times C \to \R_{\geq 0}$ instead of on $V \cup C$. The triangle inequality can then be written as $0 \leq d(v, c) \leq d(v, c') + d(v', c') + d(v', c)$ for all $v, v' \in V$ and for all $c, c' \in C$.} $V \cup C$ is \emph{consistent} with the ranking $\succ_v$ of voter $v$ if $d(v, c) \leq d(v, c')$ for all $c, c' \in C$ such that $c \succ_v c'$. 
We say that $d$ is consistent with the ranked-choice profile $\prof$ if it is consistent with the ranking $\succ_v$ for all voters $v \in V$. 
We use $\dom(\prof)$ to denote the domain of metrics consistent with $\prof$.

The \emph{(utilitarian) social cost} of a candidate $c$ with respect to a metric $d$ is defined as the candidate's sum of distances to all voters: $\cost(c, d) = \sum_{v \in V} d(v, c)$. 
A candidate $c^*_d $ is \emph{optimal} with respect to the metric $d$ if $c^*_d \in \argmin_{c \in C} \cost(c, d)$.  
The \emph{distortion} of a voting rule $f$, denoted by $\dist(f)$, is the largest possible ratio between the cost of the winner of $f$ and that of an optimal candidate $c^*_d$, with respect to the worst possible metric $d \in \dom(\prof)$. That is, 
\[\dist(f) = \max_{\prof} \sup_{d \in \dom(\prof)} \frac{\cost(f(\prof), d)}{\cost(c^*_d, d)}.\]

\subsection{Domination Graphs}

Domination graphs offer a conceptually simple approach for giving an upper bound of $3$ on the distortion of a voting rule.
Given an election $\elec = (V, C, \prof)$, the \emph{domination graph} of a candidate $c \in C$ is 
the bipartite graph $G^\elec(c) = (V, V, E_c)$ where $(v, v') \in E_c$ if and only if $c \succeq_v \top(v')$. 
The main use of these graphs is via the following lemma, due to \cite{munagala:wang:improved,DistortionDuality,gkatzelis:halpern:shah:resolving}.

\begin{lemma}
\label{lem:domination_graph}
Let $f$ be a voting rule such that for every election $\elec = (V, C, \prof)$, the domination graph $G^\elec(f(\prof))$ has a perfect matching.
Then, $f$ has distortion 3.
\end{lemma}        

Throughout the remainder of the paper, we assume that an election $\elec = (V, C, \prof)$ is given, and we~drop $\elec$ from notation when it is clear from the context.

\section{Optimal Distortion via \name} 
\label{sec:distortion}
We now introduce \name and show that it has distortion 3. 
\name requires very limited ordinal information; it only requires one each of the following two types of queries to every voter.
\begin{itemize}
\item[--] A \textit{top query} to a voter $v$ simply returns $\top(v)$. 

\item[--] A \textit{bottom-among query} to a voter $v$ regarding a subset of candidates $A$ 
returns the bottom choice of $v$ among candidates in $A$, denoted by $\bot_A(v)$.
\end{itemize} 
Notice that both can of course be easily derived when each voter's full ranking $\succ_v$ is known.

\name assigns an initial score of $\plu(c)$ to each candidate $c$; doing so only requires making a~top query to each voter.
These scores are then gradually decreased;
when the score of a candidate $c$ reaches zero, we say that $c$ is \emph{eliminated}.
Voters are processed one by one in an arbitrary order; this order can be fixed beforehand, or be adaptive and based on voters' preferences.
When a voter $v$ is processed, she decrements the score of her bottom choice among the not-yet-eliminated candidates; the bottom choice can be found by making a bottom-among query to $v$. 
The winner is the last eliminated candidate.
Pseudo-code is given as \cref{alg:plu-veto}.

\begin{algorithm}
\caption{\name}
\label{alg:plu-veto}

\enspace \ \textbf{Input:}\: An election $\elec = (V, C, \prof)$ \\
\textbf{Output:}\: A winning candidate $c \in C$

\smallskip 
\hrule

\begin{algorithmic}[1]
\smallskip

\State{\textbf{initialize} $\score(c) = \plu(c)$ for each $c \in C$}
\State{\textbf{let} $(v_1, \ldots, v_n)$ be an arbitrary ordering of $V$}

\smallskip

\For{$i=1, 2, \ldots, n$}
	\State{$A_i = \; \! \Set{c \in C}{\score(c) > 0}$}
	\State{$\, c_i \, = \; \bot_{A_i}(v_i)$} 

	\smallskip	
	
	\State{\textbf{decrement} $\score(c_i)$ \textbf{by} $1$}
\EndFor

\State{\textbf{return} $c_n$} 
\end{algorithmic}
\end{algorithm}

\begin{theorem}
\label{thm:distortion}
The distortion of \name is 3.
\end{theorem}

\begin{proof}
We show that $G(c_n)$ has a perfect matching, 
which proves that \name has distortion 3 by \cref{lem:domination_graph}.
Initially, the scores of candidates add up to $n$, 
and in each of the $n$ iterations, a positive score is decremented by $1$. 
Thus, the score of each candidate must be $0$ at the end. 
This implies that, for each candidate $c$, there are $\plu(c)$ distinct voters $v_i$ such that $c_i = c$. 
In other words, for each voter $v_i$, we can define a unique voter $v'_i$ such that $\top(v'_i) = \bot_{A_i}(v_i)$. 
This means that $v_i$ ranks any candidate in $A_i$ weakly higher than $\top(v'_i)$.
Since $\score(c_n)$ does not get to 0 until the end, note that $c_n \in A_i$ for all $i$.
Hence, each voter $v_i$ ranks $c_n$ weakly higher than $\top(v'_i)$, i.e., $(v_i, v'_i) \in E_{c_n}$.  
Thereby, we have shown that $G(c_n)$ has a perfect matching. 
\end{proof}

\cref{alg:plu-veto} is not only quite natural; it also requires only $O(\log m)$ bits of information from each voter $v_i$, namely, $\top(v_i)$ and $\bot_{A_i}(v_i)$. 
However, implementing the rule with these two queries comes with a trade-off: voters need to wait for possibly $n$ rounds after reporting their top choice.

Note that our proof of \cref{thm:distortion} also implies that there is always a candidate whose domination graph has a perfect matching.
Indeed, all attempts to resolve the optimal metric distortion conjecture, in one way or another, boiled down to proving the existence of such a candidate.   
This was done in \cite{gkatzelis:halpern:shah:resolving} by giving a~stronger existence result, called the ranking-matching lemma.
Our rule, as stated so far, does not require this stronger lemma, but in the next section, we show that a simple generalization also implies the general ranking-matching lemma.

\section{A Simple Proof of the Ranking-Matching Lemma via \fractionalveto}
\label{sec:ranking-matching}
Our proof of \cref{thm:distortion} implied that there always exists a candidate whose domination graph has a perfect matching.
The Ranking-Matching Lemma in \cite{gkatzelis:halpern:shah:resolving}  is a stronger existence result based on an extension of domination graphs in which nodes have arbitrary weights.
Given an election $\elec = (V, C, \prof)$ and weight vectors $\pvec \in \Delta(V)$ and $\qvec \in \Delta(C)$, the \emph{$(\pvec, \qvec)$-domination graph} of a candidate $\widetilde{c} \in C$ is the bipartite graph $G^{\elec}_{\pvec, \qvec}(\widetilde{c}) = (V, C, E_{\widetilde{c}}, \pvec, \qvec)$; the edge $(v, c) \in E_{\widetilde{c}}$ if and only if $\widetilde{c} \succeq_v c$.
A \emph{fractional perfect matching} of $G^{\elec}_{\pvec, \qvec}(\widetilde{c})$ is a weight function $w : E_{\widetilde{c}} \to \R_{\ge 0}$ such that  $\sum_{c \in C : (v, c) \in E_{\widetilde{c}}} w(v, c) = \p{v}$ for each voter $v$ and $\sum_{v \in V : (v, c) \in E_{\widetilde{c}}} w(v, c) = \q{c}$ for each candidate $c$.
That is, the total weight of edges incident to each node must be equal to the weight of the node.

A candidate's domination graph has a perfect matching if and only if his $(\pvec[\mathsf{uni}], \qvec[\mathsf{plu}])$-domination graph has a fractional perfect matching; here, $\p[\mathsf{uni}]{v} = 1 / n$ for all $v \in V$ and $\q[\mathsf{plu}]{c} = \plu(c) / n$ for all $c \in C$.  
The Ranking-Matching Lemma gives a stronger existence guarantee by asserting that, for any election $\elec$ and \emph{any} $\pvec \in \Delta(V)$ and $\qvec \in \Delta(C)$, there is a candidate $c \in C$ whose $(\pvec, \qvec)$-domination graph $G^{\elec}_{\pvec, \qvec}(c)$ has a fractional perfect matching.
\citet{gkatzelis:halpern:shah:resolving} give a proof by minimal counterexample for this lemma; thus, their proof is non-constructive. Moreover, it contains elaborate details for defining a minimal counterexample and intricate arguments involving smaller elections where some voters and candidates are removed.
This makes the proof somewhat harder to grasp.

We provide a much simpler and constructive proof of the Ranking-Matching Lemma.
We generalize \name in a way that allows any $\pvec \in \Delta(V)$ and $\qvec \in \Delta(C)$ to be given as input.
We refer to this voting rule as \fractionalveto, and give the pseudo-code as \cref{alg:frac-veto}.
Notice that for the special case with weights $\pvec[\mathsf{uni}]$ and $\qvec[\mathsf{plu}]$, \fractionalveto specializes to \name.

\begin{algorithm}[h]
\caption{\fractionalveto}
\label{alg:frac-veto}

\enspace \ \textbf{Input:}\: An election $\elec = (V, C, \prof)$ along with weight vectors $\pvec \in \Delta(V)$ and $\qvec \in \Delta(C)$ \\
\textbf{Output:}\: A candidate $c \in C$ whose $(\pvec, \qvec)$-domination graph has a fractional perfect matching   

\smallskip 
\hrule

\begin{algorithmic}[1]
\smallskip

\State{\textbf{initialize} $\weight(v) = \p{v}$ for each $v \in V$}
\State{\textbf{initialize} $\weight(c) = \q{c}$ for each $c \in C$}

\smallskip

\While{there is a $v \in V$ with $\weight(v) > 0$} \label{alg:while-start}
        \State{\textbf{let} $v$ be such a voter}
	\State{$A = \; \! \Set{c \in C}{\weight(c) > 0}$}
	\State{$\, c \, = \; \bot_{A}(v)$} 	
	\State{$\, \epsilon \, = \; \min \; \SET{\weight(v), \weight(c)}$}
	
	\smallskip

	\State{\textbf{decrement} $\weight(v)$ \textbf{by} $\epsilon$}
	\State{\textbf{decrement} $\weight(c)$ \textbf{by} $\epsilon$} \label{alg:while-end}

\EndWhile 

\State{\textbf{return} $c$ {\small \it (i.e., the last candidate whose weight is decremented)}}
\end{algorithmic}
\end{algorithm}

\begin{theorem}
Given any election $\elec = (V, C, \prof)$, and any weight vectors $\pvec \in \Delta(V)$ and $\qvec \in \Delta(C)$, \ \fractionalveto returns a candidate $c \in C$ whose $(\pvec,\qvec)$-domination graph $G^\elec_{\pvec,\qvec}(c)$ has a fractional perfect matching. 
\end{theorem}

\begin{proof}
We begin by observing that the \textbf{while} loop (lines \ref{alg:while-start}--\ref{alg:while-end}) terminates in at most $n+m$ iterations since in each iteration, the weight of either a voter $v$ or a candidate $c$ reaches $0$. 
Next, we show that if a candidate, say $\widetilde{c}$, wins, then $G_{\pvec,\qvec}(\widetilde{c})$ has a fractional perfect matching $w$.

First, note that, since $\widetilde{c}$ is the last candidate whose weight is decremented, $\widetilde{c} \in A$ for all iterations. 
Let us now consider an arbitrary iteration of \fractionalveto in which a voter $v$ is chosen.
By definition, $v$ ranks any candidate in $A$ weakly higher than the candidate $c = \bot_A(v)$;
in particular, $v$ ranks $\widetilde{c}$ weakly higher than $c$, i.e., $(v, c) \in E_{\widetilde{c}}$.
Let $w$ assign a weight of $\epsilon$ to the edge $(v, c)$, i.e., $w(v, c) = \epsilon$.
We now show that $w$ is a fractional perfect matching of $G_{\pvec, \qvec}(\widetilde{c})$.    

For each edge $(v, c) \in E_{\widetilde{c}}$ to which $w$ assigns positive weight, there exists an iteration where the weights of both $v$ and $c$ are decremented by $w(v, c)$.
Hence, the total weight of edges incident to each voter and candidate is equal to how much their weight is decremented until the end.
All we need to show is that the weight of each voter $v$ and candidate $c$ reaches 0 at the end as they are initialized, respectively, to $\p{v}$ and $\q{c}$. 
The total weights of voters and candidates are initially the same, and they stay so after each iteration since only a single voter's and candidate's weights are decremented, and by the same amount.  
Therefore, when the weight of each voter reaches $0$, so does each candidate's.
Thus, we have shown that $w$ is a fractional perfect matching of $G_{\pvec, \qvec}(\widetilde{c})$.     
\end{proof}

\section{A Class of Randomized Voting Rules with Distortion 3}
\label{sec:randomized}
As shown in \cref{thm:distortion}, \name has distortion 3.
Another voting rule that is well known to have distortion 3 (or $3-2/n$, to be precise) is \randomdictatorship, which returns the top choice of a 
voter chosen uniformly at random \cite{anshelevich:postl:randomized}.
In the same vein, one can view \name as choosing a voter \emph{deterministically} and returning that voter's top choice; this voter is referred to as $v'_n$ in the proof of \cref{thm:distortion}.
This suggests a more general class of randomized voting rules, interpolating between \randomdictatorship and \name: carefully rule out $k$ voters, and return the top choice of a uniformly random voter among the remaining $n-k$ voters.
In this section, we introduce such a general rule and show that it~achieves distortion at most 3 for every choice of $k \in \SET{0, 1, \ldots, n-1}$.
  
A randomized voting rule $f$ is an algorithm which, given a ranked-choice profile $\prof$, returns a probability distribution over candidates $f(\prof) \in \Delta(C)$.
Writing $\wvec = f(\prof)$, each candidate $c$ is chosen as the winner by $f$ with probability $\w{c}$. 
We refer to $\wvec$ as the \emph{winner distribution} of $f$.
The cost of a winner distribution under a given metric $d$ over $V \cup C$ is the expected cost of the winner, i.e., $\cost(\wvec, d) = \sum_{c \in C} \w{c} \cdot \cost(c, d)$.
The distortion of a randomized voting rule is still the ratio of the rule's cost to the cost of the optimum solution.

To phrase our generalized voting rule precisely, we observe that
a candidate $c$ wins under \randomdictatorship with probability proportional to his plurality score $\plu(c)$.
When the score of candidate $c_i$ is decremented at the $i^{\text{th}}$ iteration of \name, suppose that we rule out voter $v'_i$, as defined in the proof of \cref{thm:distortion}.
Since $\top(v'_i) = c_i$, the current score of candidate $c_i$ can be viewed as his plurality score with respect to the remaining voters $v'_{i+1}, \ldots, v'_n$. 

Our generalized voting rule (\cref{alg:randomized}) therefore simply runs \name for only $k < n$ iterations, then chooses a candidate with probability proportional to the residual scores.
The algorithm is formally given as \cref{alg:randomized}.

\begin{algorithm}
\caption{\randomized{k}}
\label{alg:randomized}

\enspace \ \textbf{Input:}\: An election $\elec = (V, C, \prof)$\\
\textbf{Output:}\: A winner distribution $\wvec \in \Delta(C)$

\smallskip 
\hrule

\begin{algorithmic}[1]
\smallskip

\State{\textbf{initialize} $\score(c) = \plu(c)$ for each $c \in C$}
\State{\textbf{let} $(v_1, \ldots, v_n)$ be an arbitrary ordering of $V$}

\smallskip

\For{$i=1, 2, \ldots, k$}
	\State{$A_i = \; \! \Set{c \in C}{\score(c) > 0}$}
	\State{$\, c_i \, = \; \bot_{A_i}(v_i)$} 

	\smallskip	
	
	\State{\textbf{decrement} $\score(c_i)$ \textbf{by} $1$}
\EndFor
\State{\textbf{return} $\w{c} = \score(c) / (n-k)$ for all $c \in C$} 
\end{algorithmic}
\end{algorithm}

Our main result in this section is that \randomized{k} has distortion at most 3 for any $k$.

\begin{theorem}
\label{thm:randomized}
The distortion of \randomized{k} is at most $3$ for any $k \in \SET{0,1,\ldots,n-1}$.
\end{theorem}

Our proof is based on a generalization of the flow technique from \cite{DistortionDuality} to randomized voting rules. 
It is encapsulated in \cref{lem:dual-flow} below;
this lemma is a straightforward generalization of Lemma 3.1 from \cite{DistortionDuality}. 
As in that paper, the proof is somewhat technical and long; thus, it is given in the appendix.

\medskip

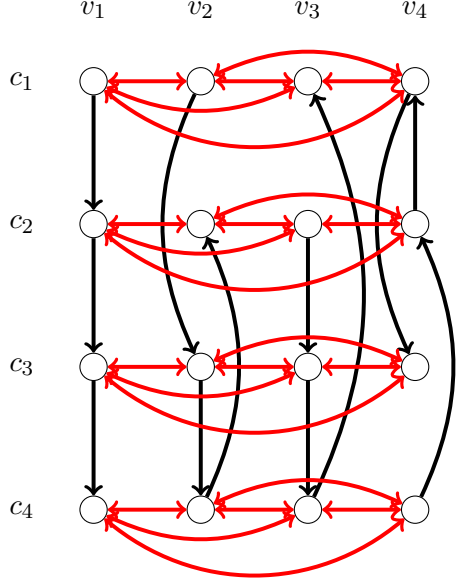
\begin{figure}[htb]
\begin{center}
\begin{tikzpicture}[auto,active/.style={circle,draw=black}, scale=0.95]
  \pgfsetxvec{\pgfpoint{0cm}{-1cm}}
  \pgfsetyvec{\pgfpoint{-1cm}{0cm}}
  
  \draw (1,4.5) node {$v_1$};
  \draw (1,3) node {$v_2$};
  \draw (1,1.5) node {$v_3$};
  \draw (1,0) node {$v_4$};

  \draw (2,5.5) node {$c_1$};
  \draw (4,5.5) node {$c_2$};
  \draw (6,5.5) node {$c_3$};
  \draw (8,5.5) node {$c_4$};

  \node[active] (v11) at (2,4.5) {};
  \node[active] (v12) at (2,3)   {};
  \node[active] (v13) at (2,1.5) {};
  \node[active] (v14) at (2,0) {};

  \node[active] (v21) at (4,4.5) {};
  \node[active] (v22) at (4,3) {};
  \node[active] (v23) at (4,1.5) {};
  \node[active] (v24) at (4,0) {};

  \node[active] (v31) at (6,4.5) {};
  \node[active] (v32) at (6,3) {};
  \node[active] (v33) at (6,1.5) {};
  \node[active] (v34) at (6,0) {};

  \node[active] (v41) at (8,4.5) {};
  \node[active] (v42) at (8,3) {};
  \node[active] (v43) at (8,1.5) {};
  \node[active] (v44) at (8,0) {};


  \draw [line width = 1.5pt,->] (v11) to (v21);
  \draw [line width = 1.5pt,->] (v21) to (v31);
  \draw [line width = 1.5pt,->] (v31) to (v41);

  \draw [line width = 1.5pt,->] (v12) to [bend right = 25] (v32);
  \draw [line width = 1.5pt,->] (v32) to (v42);
  \draw [line width = 1.5pt,->] (v42) to [bend right = 25] (v22);

  \draw [line width = 1.5pt,->] (v23) to (v33);
  \draw [line width = 1.5pt,->] (v33) to (v43);
  \draw [line width = 1.5pt,->] (v43) to [bend right = 25] (v13);

  \draw [line width = 1.5pt,->] (v44) to [bend right = 25] (v24);
  \draw [line width = 1.5pt,->] (v24) to (v14);
  \draw [line width = 1.5pt,->] (v14) to [bend right = 25] (v34);

  
  \draw [line width = 1.5pt,<->,red] (v11) to (v12);
  \draw [line width = 1.5pt,<->,red] (v12) to (v13);
  \draw [line width = 1.5pt,<->,red] (v13) to (v14);
  \draw [line width = 1.5pt,<->,red] (v11) to [bend right = 25] (v13);
  \draw [line width = 1.5pt,<->,red] (v12) to [bend left = 25] (v14);
  \draw [line width = 1.5pt,<->,red] (v11) to [bend right = 40] (v14);

  \draw [line width = 1.5pt,<->,red] (v21) to (v22);
  \draw [line width = 1.5pt,<->,red] (v22) to (v23);
  \draw [line width = 1.5pt,<->,red] (v23) to (v24);
  \draw [line width = 1.5pt,<->,red] (v21) to [bend right = 25] (v23);
  \draw [line width = 1.5pt,<->,red] (v22) to [bend left = 25] (v24);
  \draw [line width = 1.5pt,<->,red] (v21) to [bend right = 40] (v24);

  \draw [line width = 1.5pt,<->,red] (v31) to (v32);
  \draw [line width = 1.5pt,<->,red] (v32) to (v33);
  \draw [line width = 1.5pt,<->,red] (v33) to (v34);
  \draw [line width = 1.5pt,<->,red] (v31) to [bend right = 25] (v33);
  \draw [line width = 1.5pt,<->,red] (v32) to [bend left = 25] (v34);
  \draw [line width = 1.5pt,<->,red] (v31) to [bend right = 40] (v34);

  \draw [line width = 1.5pt,<->,red] (v41) to (v42);
  \draw [line width = 1.5pt,<->,red] (v42) to (v43);
  \draw [line width = 1.5pt,<->,red] (v43) to (v44);
  \draw [line width = 1.5pt,<->,red] (v41) to [bend right = 25] (v43);
  \draw [line width = 1.5pt,<->,red] (v42) to [bend left = 25] (v44);
  \draw [line width = 1.5pt,<->,red] (v41) to [bend right = 40] (v44);
\end{tikzpicture}
\end{center}
\caption{An illustration of a flow network $H_{\elec}$.
  In the example $\elec$, there are 4 voters and 4 candidates.
  The voter preferences are the following:
  $v_1: c_1 \succ c_2 \succ c_3 \succ c_4$;
  $v_2: c_1 \succ c_3 \succ c_4 \succ c_2$;
  $v_3: c_2 \succ c_3 \succ c_4 \succ c_1$;
  $v_4: c_4 \succ c_2 \succ c_1 \succ c_3$.
  Preference edges are shown in black, while sideways edges are shown in red.
  For legibility, we have omitted preference edges that could be replaced by a path of two or more other preference edges, e.g., the edge from $(v_1, c_1) \to (v_1, c_3)$.
\label{fig:flow-graph}}
\end{figure}

The key concept, adopted from \cite{DistortionDuality}, is the following flow network. Given an election $\elec$, let $H_{\elec} = (V \times C, E)$ be a directed graph with the the following edges:
\begin{itemize}[label=--]
\item For every voter $v$ and any pair of candidates $c$ and $c'$ such that $c \succ_v c'$, there is a directed \emph{preference edge} $(v,c) \to (v,c')$ in $E$.

\item For every candidate $c$ and any pair $v \neq v'$ of distinct voters, there is a (bi-directed) \emph{sideways edge} $(v,c) \leftrightarrow (v',c)$ in $E$.
\end{itemize}
An illustration of a flow network is given in \cref{fig:flow-graph}.

\medskip

For a winner distribution\footnote{\citet{DistortionDuality} considered only deterministic voting rules; in that case, the distribution $\wvec$ was restricted to have probability 1 for the deterministic winner, and 0 for all other candidates.} $\wvec \in \Delta(C)$ and a candidate $c^*$, a \emph{$(\wvec, c^*)$-flow} on $H_{\elec}$ is a circulation $\f$ (i.e., non-negative and conserving flow unless specified otherwise) in which
\begin{itemize}[label=--]
\item For each candidate $c$ and voter $v$, exactly $\w{c}$ units of flow originate at the node $(v,c)$.

\item Flow is only absorbed at nodes $(v,c^*)$.
\end{itemize}

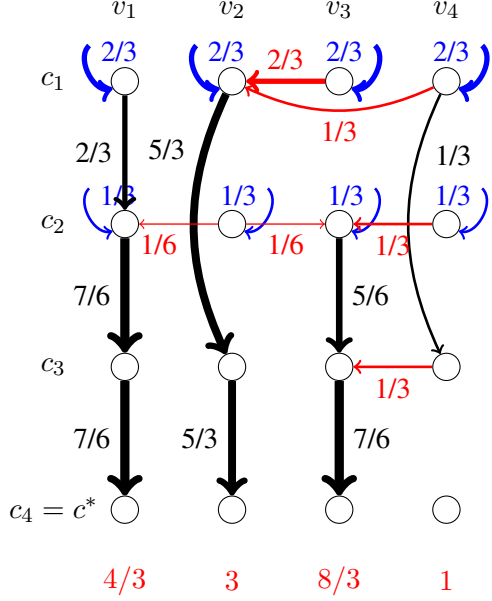
\begin{figure}[htb]
\begin{center}
  \begin{tikzpicture}[auto,active/.style={circle,draw=black}, scale=0.95]
  \pgfsetxvec{\pgfpoint{0cm}{-1cm}}
  \pgfsetyvec{\pgfpoint{-1cm}{0cm}}
  
  \draw (1,4.5) node {$v_1$};
  \draw (1,3) node {$v_2$};
  \draw (1,1.5) node {$v_3$};
  \draw (1,0) node {$v_4$};

  \draw (2,5.5) node {$c_1$};
  \draw (4,5.5) node {$c_2$};
  \draw (6,5.5) node {$c_3$};
  \draw (8,5.5) node {$c_4 = c^*$};

  \node[active] (v11) at (2,4.5) {};
  \node[active] (v12) at (2,3)   {};
  \node[active] (v13) at (2,1.5) {};
  \node[active] (v14) at (2,0) {};

  \node[active] (v21) at (4,4.5) {};
  \node[active] (v22) at (4,3) {};
  \node[active] (v23) at (4,1.5) {};
  \node[active] (v24) at (4,0) {};

  \node[active] (v31) at (6,4.5) {};
  \node[active] (v32) at (6,3) {};
  \node[active] (v33) at (6,1.5) {};
  \node[active] (v34) at (6,0) {};

  \node[active] (v41) at (8,4.5) {};
  \node[active] (v42) at (8,3) {};
  \node[active] (v43) at (8,1.5) {};
  \node[active] (v44) at (8,0) {};


  \draw [line width = 2pt,->,blue] (1.5,5) to [bend right = 75] node {2/3} (v11);
  \draw [line width = 2pt,->,blue] (1.5,3.5) to [bend right = 75] node {2/3} (v12);
  \draw [line width = 2pt,->,blue] (1.5,1) to [bend left = 75] node[swap] {2/3} (v13);
  \draw [line width = 2pt,->,blue] (1.5,-0.5) to [bend left = 75] node[swap] {2/3} (v14);

  \draw [line width = 1pt,->,blue] (3.5,5) to [bend right = 75] node {1/3} (v21);
  \draw [line width = 1pt,->,blue] (3.5,2.5) to [bend left = 75] node[swap] {1/3} (v22);
  \draw [line width = 1pt,->,blue] (3.5,1) to [bend left = 75] node[swap] {1/3} (v23);
  \draw [line width = 1pt,->,blue] (3.5,-0.5) to [bend left = 75] node[swap] {1/3} (v24);

  
  \draw [line width = 2pt,->] (v11) to node[swap] {2/3} (v21);
  \draw [line width = 3.5pt,->] (v21) to node[swap] {7/6} (v31);
  \draw [line width = 3.5pt,->] (v31) to node[swap] {7/6} (v41);

  \draw [line width = 3pt,->] (v12) to [bend right = 25] node[pos=0.3,swap] {5/3} (v32);
  \draw [line width = 3pt,->] (v32) to node[swap] {5/3} (v42);
  \draw [line width = 0.5pt,->,red] (v22) to node[pos=0.75] {1/6} (v21);
  \draw [line width = 0.5pt,->,red] (v22) to node[swap] {1/6} (v23);

  \draw [line width = 2pt,->,red] (v13) to node[swap] {2/3} (v12);
  \draw [line width = 2.5pt,->] (v23) to node {5/6} (v33);
  \draw [line width = 3.5pt,->] (v33) to node {7/6} (v43);

  \draw [line width = 1pt,->,red] (v14) to [bend left = 25] node {1/3} (v12);
  \draw [line width = 1pt,->] (v14) to [bend right = 25] node[pos=0.15] {1/3} (v34);
  \draw [line width = 1pt,->,red] (v24) to  node {1/3} (v23);
  \draw [line width = 1pt,->,red] (v34) to  node {1/3} (v33);


  \draw [red] (9,4.5) node {$4/3$};
  \draw [red] (9,3) node {$3$};
  \draw [red] (9,1.5) node {$8/3$};
  \draw [red] (9,0) node {$1$};
\end{tikzpicture}
\end{center}
\caption{An illustration of a $(\wvec, \LPC)$-flow on the flow network from \cref{fig:flow-graph},
  for $\wvec = (2/3, 1/3, 0, 0)$ and $\LPC = c_4$.
  Edges are only shown when they are used by the flow.
  Incoming flow is shown in blue.
  Flow routed along preference edges is shown in black,
  while flow routed sideways is shown in red to emphasize that it contributes to the cost.
  The amount of flow is given numerically, and also shown using the width of the lines/arcs.
  The costs for each voter are shown at the bottom of the corresponding column.
  The overall cost is the maximum cost, i.e., $3$.
\label{fig:flow-example}}
\end{figure}

The \emph{cost of $\f$ at voter $v$} is the total amount of flow absorbed at $(v,c^*)$, plus the total flow on sideways edges into or out of nodes $(v,c)$, for any candidate $c$. 
Formally, $\cost_v(\f) = \sum_{e \text{ into } (v,c^*)} \f_e + \sum_{c \neq c^*} \sum_{v' \neq v} \f_{(v',c) \leftrightarrow (v,c)}$ where $\f_{u \leftrightarrow u'} = \f_{u \to u'} + \f_{u' \to u}$.
The \emph{cost of $\f$} is $\cost(\f) = \max_{v \in V} \cost_v(\f)$.
An illustration of flows and their costs is given in \cref{fig:flow-example}.

\bigskip

The key lemma showing how to use flows to upper-bound the distortion of a voting rule is the following. 

\bigskip

\begin{lemma} \label{lem:dual-flow}
Let $f$ be a randomized voting rule, with the following property:
For every election $\elec = (V, C, \prof)$ and any candidate $c^* \in C$, on the flow network $H_\elec$, there is a $(\wvec,~c^*)$-flow $\f$ such that $\wvec = f(\prof)$ and $\cost(\f) \leq \lambda$.
Then, $\dist(f) \leq \lambda$.
\end{lemma}

We are now ready to give the proof of \cref{thm:randomized}.

\begin{extraproof}{\cref{thm:randomized}}
Fix an arbitrary candidate $c^* \in C$.
We will describe a $(\wvec, c^*)$-flow $\f$, and prove that it has cost at most 3. 
\cref{thm:randomized} then follows directly from \cref{lem:dual-flow}.

As shown in the proof of \cref{thm:distortion}, we can sort the voters as $v'_1, \ldots, v'_n$ such that $\top(v'_i) = \bot_{A_i}(v_i)$ for all $i$ in the execution of \name, i.e., without stopping early.
Define $V_k = \SET{ v_1, \ldots, v_k}$ and 
$\barV_k = \SET{v_{k+1}, \ldots, v_n}$; also 
$V'_k = \SET{v'_1, \ldots, v'_k}$ and
$\barVp_k = \SET{v'_{k+1}, \ldots, v'_n}$. 

In order to obtain a $(\wvec, c^*)$-flow, we must route flow such that for each candidate $c$ and each voter $v \in V$, exactly $\score(c)/(n-k)$ units of flow originate at the node $(v,c)$.
In particular, because $\score(c) = 0$ for candidates $c \notin A_k$, no flow originates at any nodes $(v,c)$ for $c \notin A_k$.
  
\begin{enumerate}
\item
First, consider a voter $v_i \in V_k$, so that $i \leq k$.
Since $A_k \subseteq A_i$, voter $v_i$ ranks all candidates in $A_k$ weakly higher than $\bot_{A_i}(v_i)$.
Therefore, $\f$ can route all the flow originating at nodes $(v_i, c)$ to $(v_i, \bot_{A_i}(v_i))$ along preference edges. 
At that point, there is one unit of flow at $(v_i, \bot_{A_i}(v_i))$. 
This one unit of flow is next routed to $(v'_i, \bot_{A_i}(v_i)) = (v'_i, \top(v'_i))$ using a sideways edge.
Finally, since $\top(v'_i)$ is by definition the top choice of $v'_i$, the unit can be routed to $(v'_i, c^*)$ using a preference edge, and is then absorbed.

\item 
Next, we (jointly) consider all voters $v_i \in \barV_k$.
Fix a candidate $c \in A_k$.
Because $\score(c)/(n-k)$ units of flow originate at each node $(v_i, c)$ for $v_i \in \barV_k$, and there are $|\barV_k| = n-k$ such nodes, in total, exactly $\score(c)$ units of flow originate at these nodes.
On the other hand,  $\score(c)$ is also the number of distinct voters $v'_j \in \barVp_k$ such that $\top(v'_j) = c$. 
$\f$ distributes all the $\score(c)$ units of flow from nodes $(v_i, c)$ (for $v_i \in \barV_k$) to the $\score(c)$ nodes $(v'_j, c)$ with $\top(v'_j) = c$, in a way that each voter $v'_j \in \barVp_k$ receives one unit of flow.
Then, for each voter $v'_j \in \barVp_k$ with $\top(v'_j)=c$, the flow $\f$ routes the one unit of flow from $(v'_j, c)$ to $(v'_j, c^*)$ using a preference edge; there, it is absorbed.
Because flow only originates at nodes $(v_i, c)$ with $c \in A_k$, all the flow is in fact routed to an absorbing node in this way.
\end{enumerate}

In the flow $\f$ described above, for each voter $v$, exactly one unit is sent out on sideways edges, and one unit is received on sideways edges. 
For voters in $V_k$, the unit sent is by the first case above;
for voters in $V'_k$, the unit received is by the first case. 
For voters in $\barV_k$, the unit sent is by the second case; for voters in $\barVp_k$, the unit received is by the second case. Finally, each voter $v_i$ absorbs the one received unit of flow at $(v_i, c^*)$. 
This proves that $\cost(\f) = 3$, completing the proof.
\end{extraproof}

\section{Optimal Committee Selection in Polynomial Time via \name}
\label{sec:committee}
{So far, we have focused on the case of single-winner elections.
Elections are also frequently used to choose a committee of multiple candidates.
Here, we consider a setting where a target size $k$ for the committee is given;
a \emph{committee} is then a subset of candidates $K \subseteq C$ with $|K| = k$.
A \emph{multi-winner voting rule} $f$ is an~algorithm that returns a committee $f(\prof) \subseteq C$ of the given size $k$, given a ranked-choice profile $\prof$.

The committee $K$ returned by the rule should still be ``representative'' of the set of all voters, in the same sense of being close to the voters in the metric space.
  While the distance of a voter from a single candidate is simply $d(v,c)$, many natural notions of distance from a voter to a committee (i.e., set of candidates) suggest themselves, and indeed, have been studied in the literature.
One natural notion, studied by \citet{Goel2018} and \citet{Chen2020}, is the \emph{average} distance of $v$ to the members of $C$.
However, this notion simply encourages the election of a very homogeneous committee consisting of candidates as close as possible to the median.
An alternative notion was studied by \citet{multiwinner}; this notion aims to capture the intuition that every voter should be close to at least one, or several, of the committee members.
To capture this intuition formally, \citet{multiwinner} define the \emph{$q$-cost} of a voter $v$ for a committee $K$ under a~given metric $d$ over $V \cup C$, denoted by $q$-$\cost_v(K, d)$, as the distance of $v$ from the $q^{\text{th}}$ closest candidate (to $v$) in $K$.
The \emph{$q$-social cost} of a committee $K$ is then defined as the total $q$-cost of voters for $K$, i.e., $q$-$\cost(K, d) = \sum_{v \in V} q$-$\cost_v(K, d)$.  
The $(q, k)$-distortion of a multi-winner voting rule is the ratio of the rule's $q$-social cost to the $q$-social cost of the optimum committee of size $k$, just as in \cref{sec:model:distortion}

For small values of $q$, a low $q$-cost captures the intuition that each voter should be represented on the committee (i.e., close to) one or a few candidates.
For large values of $q$, such as even $q=k$, all members of the committee must by close to most voters --- as with the sum of distances, this encourages choosing committees of candidates all of whom are close to the median of the metric space of voters, resulting in homogeneity of the committee.

\citet{multiwinner} showed an interesting trichotomy on the lowest achievable distortion for different regimes of $q$.
When $q \leq k/3$, the lowest $(q, k)$-distortion of any multi-winner voting rule can be unbounded;
when $q \in (k/3, k/2]$, the $(q,k)$-distortion of any voting rule is at least linear in the number of voters, and there exists a computationally efficient voting rule with $(q,k)$-distortion no worse than linear;
finally, when $q > k/2$, there exists a voting rule with constant distortion. 
The analysis in the third regime (large $q$) is based on the following key insight:

\begin{lemma}[Lemma 2 of \citet{multiwinner}] \label{lem:metric}
	For any election $\elec = (V, C, \prof)$, metric $d$ consistent with $\prof$ and $q > k/2$, the $q$-$\cost$ is a metric over committees of size $k$.   
\end{lemma}

Since the $q$-costs form a metric, each voter $v$'s ranking of candidates can be extended into a ranking by $v$ of all committees of size $k$.
Specifically, to decide whether voter $v$ prefers a committee $K$ over another committee $K'$, it suffices to compare her $q^{\text{th}}$ favorite candidates in $K$ and $K'$ (with ties broken arbitrarily).
Given the rankings over the committees, any single-winner rule can be used to choose the winning committee, with the same distortion guarantee by \cref{lem:metric}; \citet{multiwinner} use \pluralitymatching, achieving distortion 3.
However, since the number of committees of size $k$ is exponential in $k$, the naive approach of computing the full rankings over committees of size $k$ requires exponential running time in $k$.

\citet{multiwinner} show how to achieve polynomial running time at the cost of a loss of another factor of 3 in the distortion; i.e., they present a polynomial-time implementable voting rule with distortion no worse than 9.
They do so by proving (see Lemma~3 of \citet{multiwinner}) that there always exists a~committee composed of some voter's top $k$ choices whose distortion is no worse than three times as large as that of the optimal committee.
As a consequence of our analysis, we immediately obtain that this factor of 3 need not be lost. This is due to the following two key observations:

\begin{enumerate}
  \item In the proof of \cref{thm:distortion}, we showed that the domination graph $G(c_n)$ of the winning candidate $c_n$ has a perfect matching.
Furthermore, under \name, a candidate can only win if he~starts with a positive plurality score --- this is because a candidate is eliminated as soon as his score reaches~0.
Thus, the proof showed the existence of a candidate $c_n$ who was the first choice of at least one voter, and for whom $G(c_n)$ contains a perfect matching.

\item If $\hat{C}$ is the set of all candidates ranked first by at least one voter, then the domination graph $\hat{G}(c)$ of any candidate $c \in \hat{C}$ with respect to the candidate set $\hat{C}$ is the same as the domination graph $G(c)$ of $c$ with respect to the set of all candidates $C$.
  This is because the edge $(v,v') \in G(c)$ if and only if $c \succeq_v \top(v')$, and because $\hat{C} = \Set{c'}{c' = \top(v) \text{ for at last one voter } v}$, the top candidate $\top(v')$ is the same with respect to $C$ and $\hat{C}$.
\end{enumerate}

As a result of these two observations, a voting rule can consider just candidates in $\hat{C}$ (i.e., with at least one first-place vote), and still be assured that at least one of these candidates will have a perfect matching in $\hat{G}(c) = G(c)$, and hence achieve distortion 3.
Such a candidate $c$ can be found by running \name or \pluralitymatching on the set $\hat{C}$ of committees which are composed of the top-$k$ candidates of some voter $v$.
Because there are at most $n$ such committees (one per voter), the resulting voting rule runs in polynomial time.

\section{Conclusion and Future Work}
\label{sec:conclusion}
We showed that a simple deterministic voting rule, called \name, achieves the optimal metric distortion of 3; the proof is short and simple. We used a generalization of \name, called \fractionalveto, to prove the Ranking-Matching Lemma of \citet{gkatzelis:halpern:shah:resolving}, and we also showed that a class of randomized rules interpolating between \randomdictatorship and \name all achieve expected distortion at most 3.

\name is a very natural and potentially practical voting rule, and it would be of interest to understand which other properties it satisfies, including the standard axiomatic voting rule properties.
One difficulty is that the outcome depends on the order in which the voters are queried in the multi-round veto process.
On the other hand, this makes the structure of the set $W \subseteq C$ of potential winners (i.e., candidates who will win for at least one processing order of the voters) an interesting object of study.

Note that $W$ is a subset of candidates whose domination graph has a perfect matching, i.e., potential winners of $\pluralitymatching$.
This has several immediate implications. First, it is possible that for all orderings of voters, the selected candidate fails to be a Condorcet winner, as shown for \pluralitymatching by \citet{gkatzelis:halpern:shah:resolving}.
Second, if the metric space is $\alpha$-decisive (i.e., for each voter, the ratio between her distance to her top-ranked and second-ranked candidate is at most $\alpha$ \cite{anshelevich:postl:randomized}) then the distortion bound improves to $2+\alpha$.

Additional insights can be gleaned from the \emph{Peer Selection} setting, in which the set of voters is the same as the set of candidates, so each voter/candidate ranks herself first; as a result, all such instances are 0-decisive.
\name can be even more easily described in the Peer Selection setting: in each round $i$, the voter $v_i$ eliminates from consideration the remaining candidate furthest from her.

Here, we consider a slight variant of \name: the order of voters is chosen adaptively, and the next voter $v_{i+1}$ is always one whose first-place vote was canceled by $v_i$. In addition, the vote of $v_1$ is always canceled, so the winner is the top choice of $v_n$. A proof essentially identical\footnote{Because the vote of $v_1$ is always canceled, this rule is not technically a special case of \name.} to the one of \cref{thm:distortion} shows that this rule returns a candidate whose domination graph has a perfect matching as well.
In turn, we can use this insight to prove that there are at least \emph{two} candidates who can win in Peer Selection, i.e., $|W| \geq 2$.
First, if the process is run from an arbitrary $v_1$, then some $\hat{v} = v_n$ wins. If the process is run starting from $v_1 = \hat{v}$, the winner must be some other candidate $\tilde{v} \neq \hat{v}$, because $\hat{v}$ is eliminated in the first step. So there are at least two potential winners.

Peer selection also has interesting properties when the voters/candidates are embedded in Euclidean space $\R^D$.
Since each voter, on her turn, eliminates the voter furthest from her, the eliminated voter is always located at a corner of the convex hull of all previously uneliminated ones.
In this sense, \name ``peels away'' extreme candidates one by one\footnote{This behavior also provides some informal intuition for why the final candidate should be close to the geometric median.}:
the convex hull of the voters in $A_k$ (who are not yet eliminated after $k$ rounds) contains no voter from $\bar{A}_k$. 
A natural question is whether the set $W$ of potential winners has the same convexity property, i.e., the convex hull of $W$ contains no voter/candidate from $C \setminus W$.\footnote{For the general case (rather than Peer Selection), this is false: for example, if there is one voter and candidate each on the left and right, but only one candidate (with no voters) in the center, then the left and right candidates can win, but not the center one.} 
If true, this would show that the potential winners are in a sense ``cohesive.''

Another compelling direction concerns the incentives under \name. We described it as a~sequential process in which voters are queried about their bottom choice one by one (although of course the process can be fully simulated if each voter's full ranking is known). In describing the sequential process, we assumed that all queries are answered truthfully. While no non-trivial voting rule can be truthful in general \cite{gibbard:manipulation,satterthwaite:voting}, truthfulness can be achieved in restricted settings \cite{feldman:fiat:golomb}. An interesting direction here is to consider the ``Price of Anarchy:'' what is the worst distortion of \name if agents play a subgame perfect equilibrium in the \textsc{Veto} stage instead of truthfully revealing their bottom choice?

\bibliographystyle{plainnat}
\bibliography{davids-bibliography/names,davids-bibliography/conferences,davids-bibliography/bibliography,davids-bibliography/publications,references}

\newpage

\appendix
\section{Proof of \cref{lem:dual-flow}}
\label{sec:flow-technique}
Here, we prove \cref{lem:dual-flow}. We restate it for convenience.

\begin{rtheorem}{Lemma}{\ref{lem:dual-flow}}
Let $f$ be a randomized voting rule, with the following property:
For every election $\elec = (V, C, \prof)$ and any candidate $c^* \in C$, on the flow network $H_\elec$, there is a $(\wvec,~c^*)$-flow $\f$ such that $\wvec = f(\prof)$ and $\cost(\f) \leq \lambda$.
Then, $\dist(f) \leq \lambda$.
\end{rtheorem}

\begin{proof}
  The proof idea is exactly the same as in \cite{DistortionDuality}: we phrase an adversary's optimization problem of maximizing the expected distortion under $\wvec$ as a linear program.
  By weak duality, any feasible solution to the dual program provides an upper bound on the maximum distortion. We then show that flows directly give rise to such dual-feasible solutions.
  
  The primal linear program is directly adapted from the linear program first given in \cite{anshelevich:bhardwaj:elkind:postl:skowron,goel:krishnaswamy:munagala} in the context of a deterministic winner.
  The variables \Lpm{c}{v} of the linear program capture the distances between voters $v$ and candidates $c$.
  As such, they must be non-negative and satisfy the triangle inequality; furthermore, they have to be consistent with the voters' preferences $\succ_v$.
  The adversary's objective is to maximize the expected distortion, compared to the optimal candidate $\LPC$ with knowledge of the metric. (That is, $\LPC$ is chosen with hindsight.)
  Since the distortion is a ratio, to ensure linearity, we phrase the LP as solving the optimization problem of maximizing the \emph{expected cost} under the distribution $\wvec$, subject to the (optimum) candidate \LPC having cost exactly 1. This is equivalent, as any distances can be normalized to ensure this property.
  The resulting linear program is the following.
  
\begin{LP}[eqn:primal-lp]{Maximize}{\sum_c \w{c} \cdot \sum_v \Lpm{c}{v}}
  \Lpm{c}{v} \leq \Lpm{c}{v'} + \Lpm{c'}{v'} + \Lpm{c'}{v}
  & \mbox{ for all } c, c', v, v' 
  \qquad \qquad \qquad \qquad \mbox{ ($\triangle$ Inequality)}\\
  \Lpm{c}{v} \leq \Lpm{c'}{v}
  & \mbox{ for all } c, c', v \mbox{ such that } \Pref[v]{c}{c'} 
  \qquad \mbox{ (consistency)}\\
  \sum_v \Lpm{\LPC}{v} = 1 
  & \phantom{\mbox{ for all } c, c', v \mbox{ such that } \Pref[v]{c}{c'}} \qquad \mbox{ (normalization)}\\
  \Lpm{c}{v} \geq 0 
  & \mbox{ for all } c, v.
\end{LP}

\bigskip

After some straightforward rearrangements, the dual linear program is given as LP \eqref{eqn:dual-lp}. 
  
\begin{LP}[eqn:dual-lp]{Minimize}{\DNo}
  \DNo
  + \sum_{c': \Pref[v]{\LPC}{c'}} \DCo{v}{\LPC}{c'}
  - \sum_{c': \Pref[v]{c'}{\LPC}} \DCo{v}{c'}{\LPC}
  \\ \,  + \sum_{c', v'}
    \left(  \DTe{v}{v'}{\LPC}{c'} - \DTe{v}{v'}{c'}{\LPC}
          - \DTe{v'}{v}{\LPC}{c'} - \DTe{v'}{v}{c'}{\LPC} \right)
  \geq \w{\LPC} & \mbox{ for all } v \\
  \sum_{c': \Pref[v]{c}{c'}} \DCo{v}{c}{c'}
  - \sum_{c': \Pref[v]{c'}{c}} \DCo{v}{c'}{c}
  \\ \,  + \sum_{c', v'}
    \left(  \DTe{v}{v'}{c}{c'} - \DTe{v}{v'}{c'}{c}
          - \DTe{v'}{v}{c}{c'} - \DTe{v'}{v}{c'}{c} \right)
  \geq \w{c} & \mbox{ for all } c \neq \LPC, v \\
  \DTe{v}{v'}{c}{c'} \geq 0 &
  \mbox{ for all } v, v', c, c' \\
  \DCo{v}{c}{c'} \geq 0 & \mbox{ for all } v, c, c'.
\end{LP}

\newpage

The dual has three types of variables:
\begin{itemize}[label=--]
  \item $\DTe{v}{v'}{c}{c'}$ for the triangle inequality constraints
  \item $\DCo{v}{c}{c'}$ for the consistency constraints
  \item $\DNo$ for the normalization constraint.
\end{itemize}

Because the normalization constraint is an equality constraint, $\DNo$ is unconstrained.

Now consider a candidate $\LPC$ for whom the adversary can achieve the largest primal LP value, i.e., who is a witness for the maximum distortion.
Fix such a candidate $\LPC$ for the rest of the proof, and let $\f = \f^{\LPC}$ be a $(\wvec,\LPC)$-flow on $H_{\elec}$ of cost at most $\lambda$ --- such a flow $\f$ exists by assumption of the lemma.
We use $\f$ to determine values for the dual variables, and show that the proposed dual solution is feasible.
Furthermore, we show that the dual objective value is $\cost(\f) \leq \lambda$.
By weak LP duality, this implies that the primal is upper-bounded by $\cost(\f)$, i.e., that the adversary cannot force expected distortion larger than $\cost(\f)$ by using the candidate $\LPC$. Since $\LPC$ gave the largest distortion bound, we obtain an upper bound on the expected distortion under $\wvec$.
This will complete the proof.
The dual variables are set as follows:

\begin{itemize}[label=--]
\item For each voter $v$ and candidates $c, c'$, we set $\DCo{v}{c}{c'}$ to be the flow on the preference edge $(v,c) \to (v,c')$, i.e., we set $\DCo{v}{c}{c'} = \f_{(v,c) \to (v,c')}$.
\item For each pair of voters $v, v'$ and candidate $c$, we set $\DTe{v}{v'}{c}{\LPC}$ to be the flow on the sideways edge $(v,c) \to (v',c)$, i.e., we set $\DTe{v}{v'}{c}{\LPC} = \f_{(v,c) \to (v',c)}$.
Notice that $\DTe{v}{v'}{c}{\LPC}$ has four arguments, and we choose \LPC for the fourth argument.
\item We set $\DNo = \cost(\f) = \max_v \cost_v(\f)$.
\item All other dual variables (in particular $\DTe{v}{v'}{c}{c'}$ for $c' \neq \LPC$) are set to 0.
\end{itemize}

First observe that the dual objective value is indeed $\DNo = \cost(\f)$.
Also, non-negativity of the dual variables is obviously satisfied.
Next, we verify that both sets of dual constraints are satisfied by the dual variables values we assigned.

\begin{itemize}[label=--]
\item To verify the first set of constraints, fix a voter $v$, and rearrange the constraint to 
\begin{multline}
\w{\LPC}
  + \left( \sum_{c': \Pref[v]{c'}{\LPC}} \DCo{v}{c'}{\LPC} + \sum_{c', v'} \DTe{v'}{v}{\LPC}{c'} \right)
  - \left( \sum_{c': \Pref[v]{\LPC}{c'}} \DCo{v}{\LPC}{c'} + \sum_{c', v'} \DTe{v}{v'}{\LPC}{c'} \right)
  + \left( \sum_{c', v'} \DTe{v}{v'}{c'}{\LPC} + \sum_{c', v'} \DTe{v'}{v}{c'}{\LPC} \right)
  \\ \leq \DNo.
 \end{multline}
 We now substitute the definitions of the dual variables.
  Note that whenever $c' \neq \LPC$, our definition implies that $\DTe{v'}{v}{\LPC}{c'} = 0$.
  The left-hand side then becomes
  \begin{multline}
  \w{\LPC}
  + \left( \sum_{c': \Pref[v]{c'}{\LPC}} \f_{(v,c') \to (v,\LPC)} + \sum_{v'} \f_{(v',\LPC) \to (v,\LPC)} \right)
  - \left( \sum_{c': \Pref[v]{\LPC}{c'}} \f_{(v,\LPC) \to (v,c')} + \sum_{v'} \f_{(v,\LPC) \to (v',\LPC)} \right) \\
  + \left( \sum_{c', v'} \f_{(v,c') \to (v',c')} + \sum_{c', v'} \f_{(v',c') \to (v,c')} \right).
  \end{multline}
  Here, notice that the first term is the flow originating at $(v,\LPC)$. The second term is the total flow entering $(v,\LPC)$, and the third term is the total flow leaving $(v,\LPC)$. Thus, the combination of the first three terms is the total flow absorbed at $(v,\LPC)$.
  The final (fourth) term is the total flow along sideways edges entering or leaving nodes of the form $(v,c)$ for the particular voter $v$ fixed.

  Thus, the left-hand side of the (rearranged) dual constraint is exactly the cost $\cost_v(\f)$. In particular, it is bounded by $\cost(\f) = \DNo$, implying that the constraint is satisfied by the chosen dual variables.

\item Next, we consider the second set of dual constraints, and fix a voter $v$ and candidate $c \neq \LPC$.
We again rearrange the constraint to make the gist of the analysis clear:
\[
\w{c} 
+ \left( \sum_{c': \Pref[v]{c'}{c}} \DCo{v}{c'}{c} + \sum_{c', v'} \DTe{v'}{v}{c}{c'} \right)
- \left( \sum_{c': \Pref[v]{c}{c'}} \DCo{v}{c}{c'} + \sum_{c', v'} \DTe{v}{v'}{c}{c'} \right)
+ \left( \sum_{c', v'} \DTe{v}{v'}{c'}{c} + \sum_{c', v'} \DTe{v'}{v}{c'}{c} \right)
\leq 0.
\]
Here, we first notice that because $c \neq \LPC$, by definition, all the terms $\DTe{v}{v'}{c'}{c}$ and $\DTe{v'}{v}{c'}{c}$ in the last two sums are 0.
Similarly, in the second and fourth sums, all terms for $c' \neq \LPC$ are 0.
Substituting the definitions for the remaining dual variables, the left-hand side equals
\[
\w{c} 
+ \left( \sum_{c': \Pref[v]{c'}{c}}   \f_{(v,c') \to (v,c)} + \sum_{v'} \f_{(v',c) \to (v,c)} \right)
- \left( \sum_{c': \Pref[v]{c}{c'}}   \f_{(v,c) \to (v,c')} + \sum_{v'} \f_{(v,c) \to (v',c)} \right).
\]
Here, observe that the first term $\w{c}$ is the amount of flow inserted at $(v,c)$, the second term is the amount of flow entering the node $(v,c)$ along preference or sideways edges, and the third term is the amount of flow leaving the node $(v,c)$ along preference or sideways edges.
Thus, the left-hand side is exactly the difference between incoming and outgoing flow at $(v,c)$, and because $\f$ was assumed to be a valid $(\wvec,\LPC)$-flow on $H_{\elec}$ (and $c \neq \LPC$), this net flow must be 0. Therefore, the dual constraint is satisfied.
\end{itemize}

  This completes the proof.
\end{proof}

\end{document}